\documentclass[conference]{IEEEtran}
\IEEEoverridecommandlockouts
\usepackage[table, dvipsnames]{xcolor}
\usepackage{cite}
\usepackage{amsmath,amssymb,amsfonts}
\usepackage{algorithmic}
\usepackage{graphicx}
\usepackage{textcomp}
\usepackage{xcolor}
\def\BibTeX{{\rm B\kern-.05em{\sc i\kern-.025em b}\kern-.08em
    T\kern-.1667em\lower.7ex\hbox{E}\kern-.125emX}}
\usepackage{comment}
\begin{document}


\title{Optimal Transport-based Semantic Alignment for LLM-based Audio-Visual Speech Recognition*\\	
}
\author{\IEEEauthorblockN{Xugang Lu$^1$, Peng Shen$^1$, Yu Tsao$^2$, Hisashi Kawai$^1$}
\IEEEauthorblockA{\textit{1. National Institute of Information and Communications Technology, Kyoto, Japan} \\
\textit{2. Research Center for Information Technology Innovation, Academia Sinica, Taiwan} 
}
}


\maketitle

\begin{abstract} 
 Large language model (LLM)-based audio-visual speech recognition (LLM-AVSR) has recently demonstrated strong robustness in adverse acoustic environments by leveraging complementary audio and visual information. Existing approaches typically employ independently pretrained acoustic and visual encoders, whose outputs are projected and fused as soft prompts to condition an LLM for speech recognition. However, most methods perform multimodal fusion without explicitly addressing the representational discrepancy between audio, visual and text modalities, potentially limiting the effectiveness of cross-modal integration. In this paper, we propose an optimal transport (OT)-based semantic alignment framework for LLM-AVSR. The proposed method explicitly bridges the modality gap by aligning the acoustic and visual representations with reference to the linguistic embedding space of the LLM before multimodal fusion. Specifically, OT is used to estimate probabilistic coupling matrices that characterize structured correspondences between modality-specific features and linguistic embeddings. The resulting OT couplings are further utilized as soft pseudo-labels to supervise contrastive learning, encouraging the extraction of semantically coherent and cross-modal consistent audio-visual representations. By anchoring multimodal features to the linguistic space of the LLM, the proposed framework facilitates more effective multimodal fusion and decoding. We implement the proposed framework using a Whisper-based acoustic encoder, an AV-HuBERT-based visual encoder, and a LLaMA3.2-3B decoder. Experiments conducted on the LRS3-TED benchmark demonstrate consistent improvements over strong baselines and achieve state-of-the-art performance under both clean and noisy evaluation conditions across a wide range of signal-to-noise ratios (SNRs). 
\end{abstract}


\section{Introduction}
Audio-visual speech recognition (AVSR) has attracted increasing attention due to its robustness in adverse acoustic environments. By leveraging complementary visual cues, such as modality-invariant features \cite{MIRGAN2023}, modality temporal dynamics \cite{VTDynamic2024}, and lip movements \cite{MaICASSP2021}, AVSR systems can substantially alleviate performance degradation caused by background noise \cite{MaICASSP2021, MaICASSP2023, CappellazzoICASSP2025, VSPLLM2024}. Recent advances in AVSR have shown that the effective integration of audio and visual information within Conformer-based encoder-decoder architectures, often combined with hybrid CTC/Attention objectives \cite{CTCAED}, can significantly improve recognition performance \cite{MaICASSP2021, MaICASSP2023}.

With the rapid development of large language models (LLMs), recent studies have extended AVSR to LLM-based speech recognition frameworks (LLM-AVSR) \cite{WFlamingo2024, CappellazzoICASSP2025, CappellazzoOmni2025, MMSLLAMA2025}. These systems typically employ pretrained modality-specific encoders to extract acoustic and visual representations, which are subsequently projected into the embedding space of an LLM and used as soft prompts for autoregressive transcription generation. Benefiting from the strong contextual modeling and linguistic reasoning capabilities of LLMs, such approaches have demonstrated improved robustness and generalization.

However, visual speech recognition (VSR) inherently suffers from ambiguity because multiple phonemes may correspond to identical or highly similar lip shapes \cite{VALLR2025, VSPLLM2024}. This ambiguity becomes even more severe in continuous speech due to co-articulation effects, where the same viseme may correspond to different phonetic realizations depending on the surrounding context \cite{VALLR2025, VSPLLM2024}. Although acoustic information can partially compensate for this limitation, speech signals themselves remain vulnerable to environmental noise and distortion. In this context, the strong linguistic priors learned by LLMs provide an effective mechanism to resolve ambiguities in both audio and visual modalities during decoding \cite{WFlamingo2024, MMSLLAMA2025, CappellazzoICASSP2025}. Moreover, the flexibility of LLMs enables the incorporation of additional linguistic knowledge, facilitating multilingual AVSR even when paired audio-visual training data are limited \cite{ZeroAVSR2025}.

Despite their promising performance, most existing LLM-AVSR systems directly fuse acoustic and visual representations without explicitly addressing the representational discrepancy between modalities. Since acoustic and visual encoders are pretrained independently using different objectives and data distributions, their latent feature spaces often exhibit substantial structural mismatch. Consequently, simple projection and concatenation strategies may fail to fully exploit the complementary information contained in the two modalities. Furthermore, when multimodal representations are projected into the LLM embedding space, the discrepancy between modality-specific features and linguistic token embeddings may hinder effective semantic grounding and limit the ability of the LLM to utilize multimodal information. Another important challenge is that temporal synchronization does not necessarily imply semantic synchronization. Although audio and visual streams are naturally aligned in time, their contributions to linguistic content may occur at different temporal locations \cite{Unsynchronized2025}. Therefore, directly combining temporally aligned features may not effectively reduce ambiguity. Instead, multimodal representations should be aligned according to their linguistic relevance, enabling the extraction of jointly informative features that better correspond to linguistic units. This semantic alignment improves cross-modal consistency and facilitates more effective multimodal fusion \cite{AlignFuse2021}.

To address the problem of feature and distribution alignment, optimal transport (OT) has been widely adopted in machine learning, including domain adaptation and transfer learning \cite{Courty2014}, natural language processing \cite{KusnerICML2015, Yokoi2020}, linguistic knowledge distillation for ASR \cite{LuASRU2023, LuICASSP2024}, and audio-visual captioning tasks \cite{LAVCAP2025}. OT provides a principled framework for establishing geometry-aware correspondences between heterogeneous feature distributions \cite{PeyreBook2019}. In this paper, we propose an OT-based semantic alignment framework that explicitly bridges the modality gap before multimodal fusion in LLM-AVSR. In the proposed framework, acoustic features, visual features, and linguistic token embeddings are modeled as empirical probability measures in a shared latent space. Specifically, hidden representations generated by the Whisper acoustic encoder and the AV-HuBERT visual encoder are treated as source distributions, while token embeddings of the target text sequence from the LLM serve as reference linguistic distributions. OT is then employed to compute soft coupling matrices that characterize structured transport relationships between modality-specific representations and linguistic embeddings. Using these couplings, the model learns audio-visual representations that are semantically aligned with the linguistic space of the LLM rather than relying solely on temporal correspondence. Furthermore, the OT-derived couplings are utilized as soft pseudo-labels to supervise contrastive learning, encouraging the extraction of linguistically relevant and cross-modally consistent feature representations. As a result, the learned multimodal embeddings become better aligned with one another and with the downstream token embedding space, facilitating more effective conditioning of the LLM for speech recognition.

We implement the proposed framework using a Whisper-based acoustic encoder, an AV-HuBERT-based visual encoder, and the LLaMA3.2-3B language model as the decoder. The proposed OT alignment module is inserted between the modality-specific encoders and the projection layers that interface with the LLM, requiring no modification to the pretrained encoders or the LLM architecture. The entire system is optimized using the ASR objective and evaluated on the LRS3-TED benchmark. Experimental results demonstrate that the proposed method consistently improves recognition performance over strong baselines and achieves state-of-the-art results on the LRS3-TED dataset. The main contributions of this work are summarized as follows: (1) We propose an OT-based semantic alignment framework that aligns multimodal representations with the linguistic embedding space of the LLM prior to fusion. (2) We introduce OT-derived probabilistic couplings as soft supervision for contrastive learning, promoting semantically coherent and cross-modally consistent feature representations. (3) Experiments with the LRS3-TED benchmark demonstrate the effectiveness of the proposed framework and establish state-of-the-art performance.

\section{Proposed method}
\begin{figure*}[tbh]
	\centering
	\includegraphics[width=15cm, height=9cm]{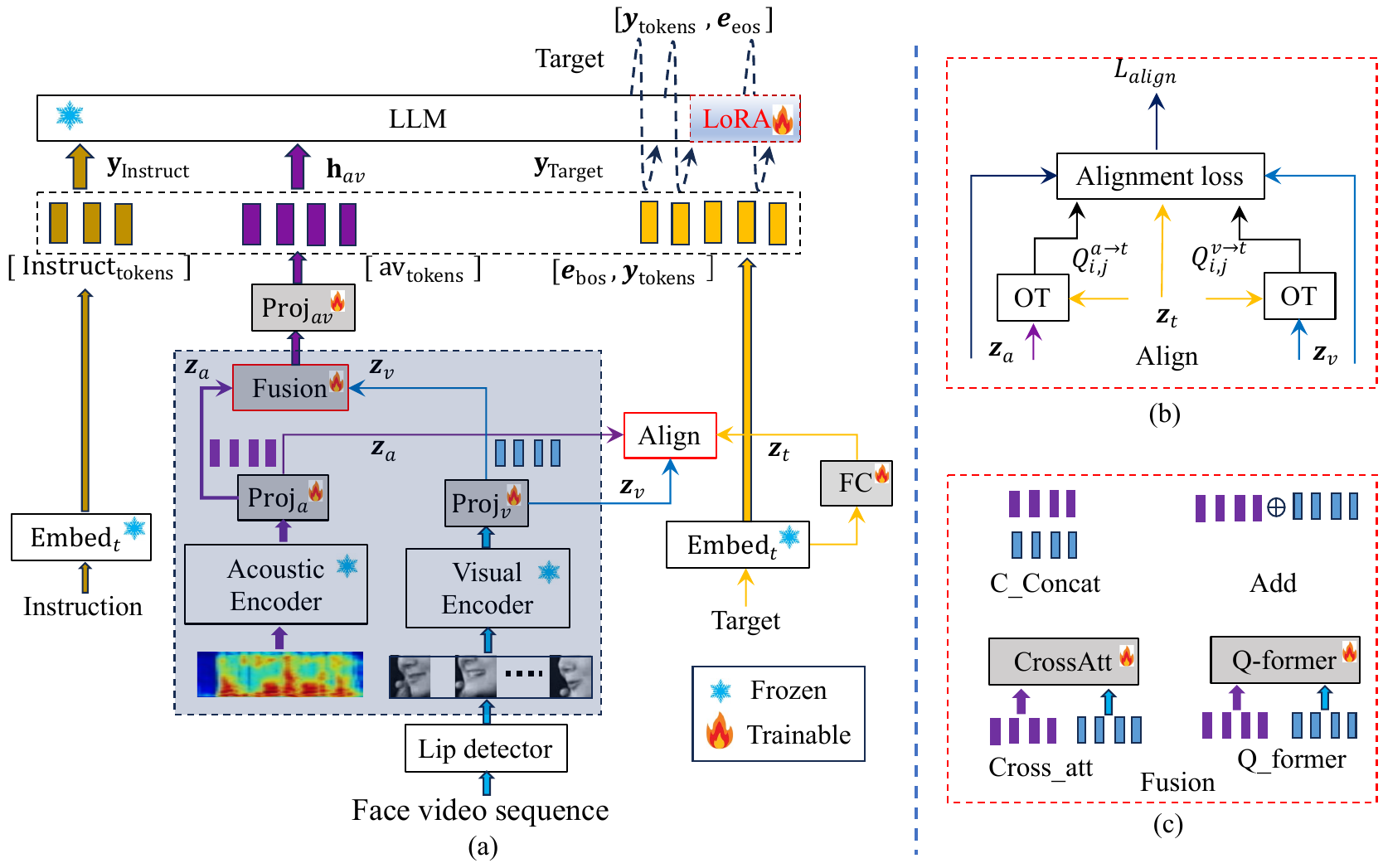}
	\caption{The proposed LLM-based audio-visual ASR (LLM-AVSR) model framework: (a) Model architecture, (b) OT-based alignment module, (c) Modules for several alternative fusion methods.}		
	\label{fig:fig1}
\end{figure*}

The proposed model framework is shown in Fig. \ref{fig:fig1}. In this figure, the inputs are composed of four components (in the training stage), i.e., the instruction text, the acoustic signal, the visual signal, and the target text. Both the instruction and the target text are processed with token embedding (with the ${\rm Embed}_t$ block). Based on the process, the instruction token embedding and target text embedding are obtained. The audio signal is processed with an acoustic encoder while the visual signal is processed via an visual encoder, and is integrated as input to an LLM decoder. The details are explained in the following.     

\subsection{Multimodal feature encoding}
For the acoustic encoder, several alternatives can be adopted, e.g., Whisper model \cite{Whisper2023}, wav2vec2.0 \cite{Wav2Vec2.0}, HuBERT \cite{HuBERT}, etc. While wav2vec2.0 and HuBERT are based on unlabeled audio data with self-supervised training, Whisper model is based on supervised training optimized for ASR. For video or image encoding, the AV-HuBERT model \cite{AVHUBERT}, the Auto-AVSR model \cite{MaICASSP2023} are widely used. Regarding LLMs, LLaMA \cite{LLaMA3.2}, Qwen \cite{Qwen}, etc., are widely adopted as decoders for down-stream tasks. Before encoded features are input to LLM, linear projections are applied to bridge the modality gaps and to keep LLM to be used without large changes. Although the choice of these encoders and LLM-based decoders could affect the final performance a lot, due to popularity and consistency with others' work, in our study, Whisper acoustic encoder for its target as ASR, AV-HuBERT for visual encoder since many studies showed that AV-HuBERT encoder could encode viseme-like information to help discriminate phonemes for ASR \cite{AVHUBERT}, and LLaMA for linguistic decoder with auto-regressive generation of ASR transcriptions. The feature process in each module is formulated as follows:
\begin{equation}
	\begin{array}{l}
		\begin{aligned}
			{\bf z}_a  &= {\rm Proj}_a \left( {{\rm Encoder}_a \left( {{\bf x}_a} \right)} \right) \\ 
			{\bf z}_v  &= {\rm Proj}_v \left( {{\rm Encoder}_v \left( {{\bf x}}_v \right)} \right) \\ 
			{\bf y}_{{\rm Instruct}}  &= {\rm Embed}_t \left( {{\rm Instruct}} \right) \\ 
			{\bf y}_{{\rm Target}}  &= {\rm Embed}_t \left( {{\rm Target}} \right) \\ 
			{\bf z}_t  &= {\rm FC}\left( {{\bf y}_{{\rm Target}} } \right) \\ 
		\end{aligned}
	\end{array}
\end{equation}
In these formulations, ${{\bf x}_a}$ and ${{\bf x}_v}$ are acoustic and visual input vectors, while ${{\rm Encoder}_a}(.)$ and ${{\rm Encoder}_v}(.)$ are acoustic and visual encoders, respectively, ${\rm Embed}_t(.)$ is text token embedding which is the same as used in the LLM, ${\rm Proj}_a(.)$ and ${\rm Proj}_v(.)$ are linear projection or connector layers for audio and visual features, ${\rm FC(.)}$ is a linear full connection layer for dimension conversion. After these processing, we obtain features as ${\bf Z}_a  \in R^{L_a  \times d} ,{\bf Z}_v  \in R^{L_v  \times d} ,{\bf Z}_t  \in R^{L_t  \times d} $, $L_a  = L_v  = L$ (length of temporal sequence). ${\bf y}_{{\rm Instruct}}$ is used directly as input to LLM, while the target embedding ${\bf y}_{{\rm Target}}$ is projected with a linear transform for dimension conversion as ${\bf z}_t$ which will be used as reference for the subsequent modality alignment and fusion process. In our study, both the acoustic and visual encoders are frozen while the LLM is fine tuned with the LoRA algorithm \cite{LoRA}.
\subsection{Feature alignment based on OT}
In order to align between different modality feature sequences, we regard the acoustic, visual and linguistic feature sequences as empirical probability distributions with $\mu_a$, $\mu_v$, and $\mu_t$. Optimal transport between audio and text embedding distributions is defined as (with entropy regularization of OT coupling):
\begin{equation}
	\mathop {\max }\limits_{Q^{a \to t}  \in \prod {\left\{ {\mu_a ,\mu_t } \right\}} } \left\langle {Q^{a \to t} ,S^{a \to t} } \right\rangle  + \lambda ^a H\left( {Q^{a \to t} } \right),
	\label{eq:Qa2t}
\end{equation}
while the OT correspondence between visual and text embedding distributions is defined as:
\begin{equation}
	\mathop {\max }\limits_{Q^{v \to t}  \in \prod {\left\{ {\mu_v ,\mu_t } \right\}} } \left\langle {Q^{v \to t} ,S^{v \to t} } \right\rangle  + \lambda ^v H\left( {Q^{v \to t} } \right),
	\label{eq:Qv2t}
\end{equation}
where ${Q^{a \to t} } \!\in \! R_ + ^{L  \times L_t }$ and ${S^{a \to t} }$ are transport coupling and similarity matrices between audio and text sequences, ${Q^{v \to t} } \!\in \! R_ + ^{L  \times L_t } $ and ${S^{v \to t} }$ are transport coupling and similarity matrices between visual and text sequences, respectively. $H\left( Q \right) \!=\!  - \sum\limits_{i,j} {Q_{i,j} \log Q_{i,j} } $ is the entropy of the transport coupling matrix defined in each audio-text and visual-text modalities. The similarity matrices ${S^{a \to t} }$ and ${S^{v \to t} }$ are estimated from the cosine similarity function for each element as:
\begin{equation}
	\begin{array}{l}
		s^{a \to t}  = \frac{{\left\langle {{\bf z}_a ,{\bf z}_t } \right\rangle }}{{\left\| {{\bf z}_a } \right\|.\left\| {{\bf z}_t } \right\|}} \\ 
		s^{v \to t}  = \frac{{\left\langle {{\bf z}_v ,{\bf z}_t } \right\rangle }}{{\left\| {{\bf z}_v } \right\|.\left\| {{\bf z}_t } \right\|}} \\ 
	\end{array}
\end{equation}
The coupling matrices on the joint probability distributions are defined as follows:
\begin{equation}
	\prod {\left\{ {\mu_a ,\mu_t } \right\}} = \left\{ {Q^{a \to t} {\bf 1}_{L_t }  = \mathbf{p}^a ,(Q^{a \to t} )^T {\bf 1}_{L_a }  = \mathbf{p}^t } \right\}, 
\end{equation}
and
\begin{equation}
	\prod {\left\{ {\mu_v ,\mu_t } \right\}} = \left\{ {Q^{v \to t} {\bf 1}_{L_t }  = \mathbf{p}^v ,(Q^{v \to t} )^T {\bf 1}_{L_v }  = \mathbf{p}^t } \right\},
\end{equation}  
where ${\bf 1}_{L_t } \! \in \! R^{L_t }$ and ${\bf 1}_{L_a } \!\in \! R^{L_a }$ are vectors of ones, ${\bf p}^a$, ${\bf p}^v$ and ${\bf p}^t$ are vectors that correspond to the three distributions with $\mu _a \left( {{\bf p}^a ,{\bf Z}_a } \right) \mathop  = \limits^\Delta \sum\limits_{i = 1}^{L_a } {p_i^a \delta \left( {{\bf z}_a^i } \right)} $, $\mu _v \left( {{\bf p}^v ,{\bf Z}_v } \right) \mathop  = \limits^\Delta \sum\limits_{i = 1}^{L_v } {p_i^v \delta \left( {{\bf z}_v^i } \right)} $ and $\mu _t \left( {{\bf p}^t ,{\bf Z}_t } \right) \mathop  = \limits^\Delta \sum\limits_{i = 1}^{L_t } {p_i^t \delta \left( {{\bf z}_t^i } \right)} $, respectively. 

This entropy-regularized OT can be solved based on a fast approximation using iterative Sinkhorn-type algorithms \cite{Cuturi2013}. For the OT problem between two distributions $\mu_1$ and $\mu_2$, the OT coupling could be obtained as:
\begin{equation}
	{Q}^* = \operatorname{diag}(\mathbf{u}) K \operatorname{diag}(\mathbf{v}), \quad K_{i,j} = \exp\left( - \frac{S_{i,j}}{\lambda} \right),
\end{equation}
where \( \mathbf{u} \in \mathbb{R}^m \) and \( \mathbf{v} \in \mathbb{R}^n \) are the scaling vectors, $S$ is a similarity matrix with element $S_{i, j}$ defined in the samples of the two distributions, and $K$ is the Gibbs kernel matrix with elements defined as $K_{i,j}$.
The iteration process to update the scaling factors to solve OT is obtained as \cite{Cuturi2013, PeyreBook2019}:
\begin{equation}
		\mathbf{u}^{(k+1)} =  \frac{\mathbf{p}^1}{K \mathbf{v}^{(k)}} , \quad
		\mathbf{v}^{(k+1)} = \frac{\mathbf{p}^2}{K^\top \mathbf{u}^{(k+1)}},
	\label{eq:scaling}
\end{equation}
where ${\mathbf {p}^1} = \left[ {p_1^1 ,p_2^1 ,...,p_{L_1 }^1 } \right]$ and ${\mathbf {p}^2} = \left[ {p_1^2 ,p_2^2 ,...,p_{L_2 }^2 } \right]$ are the probability weighting coefficients for samples from the two distributions $\mu_1$ and $\mu_2$, and with size of ${L_1 }$ and ${L_2}$. These updates are repeated until convergence or stopped when reaching a given threshold. 

\subsection{Aligning non-informative modality features with virtual buckets}
\begin{figure}[tbh]
	\centering
	\includegraphics[width=6cm, height=3.5cm]{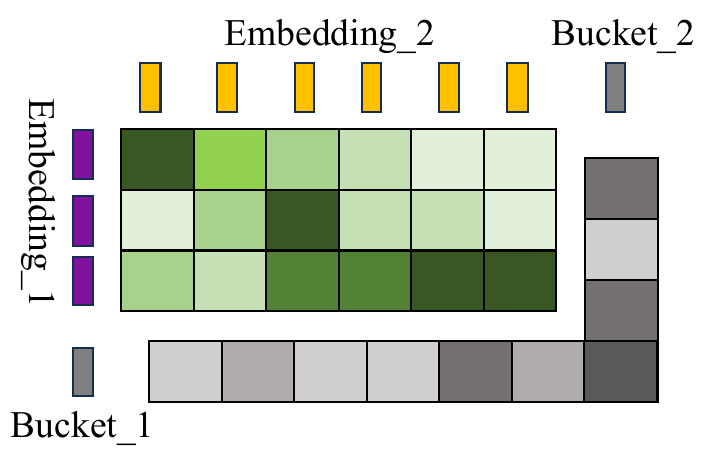}
	\caption{Virtual buckets for aligning non-informative features.}		
	\label{fig:fig2}
\end{figure}

In assigning mass of feature distributions between two modalities, it is possible that non-informative features (either noise frames or background outliers) should not be matched. For example, frame segments of acoustic or video clips, are sometimes irrelevant to linguistic transcriptions which should not be aligned in matching. Inspired by the work in \cite{SuperGlue2020, AlignWild2022}, we introduce virtual buckets in OT based matching. The process is illustrated in Fig. \ref{fig:fig2}, where the gray positions correspond to the OT coupling for non-informative features. The corresponding similarity matrix is constructed as follows:  
\begin{equation}
	\begin{array}{l}
		\tilde S_{i,j}  = S_{i,j} ,\quad i \le m,j \le n \\ 
		\tilde S_{m + 1,j}  = \tilde S_{i,n + 1}  = \tilde S_{m + 1,n + 1}  = \tilde s, \\ 
	\end{array}
	\label{eq:bucket}
\end{equation} 
where $\tilde{s}$ is a similarity margin to determine the frames with very small similarities that do not occur in semantic alignment. 
Executing OT based on this augmented similarity matrix, we could obtain the OT coupling $\tilde Q$, and keep $Q= \tilde Q_{[:m,:n]}$. The basic intuition behind it is to filter out those noise or outlier frames during transport assignment either between audio and text, or between video and text matching. 

\subsection{Alignment loss in feature learning}
The obtained OT coupling matrices from Eqs. (\ref{eq:Qa2t}) and (\ref{eq:Qv2t}) provide probabilistic correspondence for audio-text and video-text pairs, respectively. These correspondence can be used as soft pseudo-labels to supervise the feature contrastive learning defined as cross-entropy loss \cite{ContrastiveCluster2020}:
\begin{equation}
	L_{\rm align}  =  - \sum\limits_i {\sum\limits_j {\left( {Q_{i,j}^{a \to t } \log P_{i,j}^{a \to t}  + Q_{i,j}^{v \to t } \log P_{i,j}^{v \to t} } \right)} },
\end{equation}
where 
\begin{equation}
	P_{i,j}^{a \to t}  = \frac{{\exp \left( {{{S_{i,j}^{a \to t } } \mathord{\left/
						{\vphantom {{S_{i,j}^{a \to t } } \tau }} \right.
						\kern-\nulldelimiterspace} \tau }} \right)}}{{\sum\limits_k {\exp \left( {{{S_{i,k}^{a \to t } } \mathord{\left/
							{\vphantom {{S_{i,j}^{a \to t } } \tau }} \right.
							\kern-\nulldelimiterspace} \tau }} \right)} }},\quad P_{i,j}^{v \to t}  = \frac{{\exp \left( {{{S_{i,j}^{v \to t } } \mathord{\left/
						{\vphantom {{S_{i,j}^{^{v \to t} } } \tau }} \right.
						\kern-\nulldelimiterspace} \tau }} \right)}}{{\sum\limits_k {\exp \left( {{{S_{i,k}^{v \to t } } \mathord{\left/
							{\vphantom {{S_{i,j}^{v \to t } } \tau }} \right.
							\kern-\nulldelimiterspace} \tau }} \right)} }},
						\label{eq:tempscale}
\end{equation}
with $\tau$ as an adjustable temperature scale parameter. This alignment loss will be added to the final learning objective function during model training.
\subsection{Multimodal feature fusion}
As shown in Fig. (\ref{fig:fig1}-c), several alternative fusion strategies could be applied. In our study, four feature fusion methods are tested, i.e., channel-wise concatenation, addition, and multi-head attention cross-attention \cite{CMAlign2023}, as well as Q-former \cite{QFormer}-based fusion which integrate audio and visual information as input to LLM, and are denoted as C\_Concat, Add, Cross\_att and Q\_former in Fig. (\ref{fig:fig1}-c), respectively. Their formulations are as follows.
\begin{equation}
	\begin{array}{l}
		\begin{aligned}
			{\bf h}_{\rm{{C}\_Concat}}  &= \left[ \begin{array}{l}
				{\bf z}_a  \\ 
				{\bf z}_v  \\ 
			\end{array} \right] \in R^{L \times 2d}  \\ 
			{\bf h}_{\rm{Add}}  &= {\bf z}_a  + {\bf z}_v  \in R^{L \times d}  \\ 
			{\bf h}_{\rm{{Cross}\_att}}  &= {\rm CrossAtt}\left( {{\bf z}_a ,{\bf z}_v, {\bf W}_{att} } \right) \in R^{L \times d}  \\ 
			{\bf h}_{\rm{{Q}\_former}}  &= {\rm QFormer}\left( {\left[ \begin{array}{l}
					{\bf z}_a  \\ 
					{\bf z}_v  \\ 
				\end{array} \right] ,{\bf W}_q } \right) \in R^{L_q  \times d_q }  \\ 
		\end{aligned}
	\end{array}			
	\label{eq:fusion}
\end{equation}
where ${\bf h}_{\rm{{C}\_Concat}}$, ${\bf h}_{\rm{Add}}$, ${\bf h}_{\rm{{Cross}\_att}}$, ${\bf h}_{\rm{{Q}\_former}}$ are channel-concatenation, addition, cross-attention, and Q-former-based modality fusion methods, ${\bf W}_{att} $ and ${\bf W}_q $ are learnable cross-attention and Q-former model parameters, respectively, $L_q$ is querying length which is proportional to the duration of input clip length. Although temporal-concatenation, i.e., concatenating audio and visual token sequences along the temporal axis, could also be used \cite{CappellazzoICASSP2025}, the length of token sequence used as input to LLMs will be increased resulting in large model complexity, while our concatenation is along the channel dimension without increasing the input token sequence length to LLMs. There are no theoretical advantages among these alternatives, in our study, we examined these fusion strategies, and finally empirically choose the feature channel concatenation as our baseline fusion method (results will be given in experiments). The fused feature is projected to the LLM space as:
\begin{equation}
	{\bf h}_{av}  = {\rm Proj}_{av} \left( {{\bf h}_{\rm fusion} } \right),  
\end{equation}
where ${\rm Proj}_{av}(.)$ is a linear projection layer for the audio-visual fused feature, and its input ${\bf h}_{\rm fusion}$ is chosen from the alternative fusion methods in Eq. (\ref{eq:fusion}).
\subsection{Objective function}
In training stage, each training data sample is a triple element corresponding to paired audio, visual and text, the loss is defined as combination of LLM based auto-regressive (AR) cross entropy loss and alignment loss:
\begin{equation}
	L_{\rm Total}  = L_{\rm AR}  + \alpha L_{\rm align}.
	\label{eq:TotalLoss}
\end{equation}
The loss $L_{\rm AR}$ is defined as:
\begin{equation}
	L_{\rm AR}  =  - \frac{1}{N}\sum\limits_i {\log \left( {y_i |{\bf y}_{{\rm Target} < i} ,{\bf h}_{av} ,{\bf y}_{{\rm Instruct}} } \right)}  
\end{equation}
where ${\bf y}_{{\rm Target} < i}$ means target labels before the $i$-th output, $\alpha$ is the weighting coefficient. In the inference stage, only audio and visual signals are used as shown in Fig. \ref{fig:fig1}. 

\section{Experiments and results}
\label{sect:experiments}
We conduct AVSR experiments on the LRS3-TED corpus, a large-scale benchmark data set of clips that are collected from TED talks \cite{LRS3}. This data set contains thousands of hours of unconstrained, in-the-wild speech videos with significant speaker, pose, illumination, and background variability, making it a challenging benchmark for robust AVSR evaluation. The videos of the original LRS3-TED data have a resolution of 224$\times$224 pixels and a frame rate of 25 fps. With face detector and lip region crop pre-processing, visual inputs consisting of cropped mouth regions at 96$\times$96 pixels are obtained \cite{MaICASSP2021, MaICASSP2023}. We follow the standard train, validation and test split provided by the data set and use the official test set for performance comparison, for more specific, 433 hours of audio-visual data are used for model training (163374 utterances), 1 hour held-out utterances for validation set (1200 utterances), and 1 hour data for test (1321 utterances). 

In the training stage, cropping visual patches size of 88$\times$88 with random cropping, horizontal flipping are used for visual feature extraction. These cropped lip regions are used as inputs to the pretrained AV-HuBERT encoder (large size) model for lip-region visual embeddings \cite{MaICASSP2023}, and these visual speech representations encode cues of lip movements and articulatory dynamics that are supposed to provide complementary information to the acoustic modality, particularly in noisy environments where speech signals may be corrupted. For acoustic input, 16 kHz audio signals are fed into a pretrained Whisper encoder (medium size) to extract frame-level acoustic representations. In addition to the downsample process in the Whisper model, another temporal convolution layer with a stride of 2 was applied which made the sampling rate of acoustic embeddings fit to those of visual embeddings. Moreover, audio data augmentation is applied in noisy condition training to improve robustness. The modality-specific features are then projected into the LLM embedding dimension and fused as soft prompts for ASR decoding with an LLM (the LLaMA3.2-3B is applied in this paper). The LLM is responsible for modeling long-range contextual dependencies and generating the final transcription autoregressively conditioned on the fused audio and visual information. For comparison, we followed the same parameter settings for hidden feature dimensions and model optimization strategy as in \cite{CappellazzoICASSP2025, MMSLLAMA2025}, where LoRA rank of $16$, scaling factor of $32$ for LLM fine-tuning, Adam optimizer with initial learning rate $1e{\text-}4$, $5$k warm-up steps in a total of $30$k training steps. 

In evaluation, in addition to clean evaluation, we also evaluate noisy test data. The noisy test data are simulated with additive noise at multiple signal-to-noise ratio (SNR) levels (-5dB, 0dB, 5 dB) to assess its robustness. Performance is measured using the word error rate (WER).
\subsection{Effect of different fusion methods}
Different fusion methods represent different integration strategies with LLM. As shown in Fig. \ref{fig:fig1}, four fusion methods are examined in clean and noisy training conditions. Moreover, temporal-wise concatenation-based fusion is also examined in our experiments although it is not shown in Fig. \ref{fig:fig1}-c. Before testing on our proposed method, we built baseline systems in two training conditions, one is for clean training condition, i.e., the input of speech signal is clean speech without data augmentation. In noisy audio training conditions, the input of audio speech is with data augmentation (with a 75\% probability of noisy speech) where babble noise from NOISEX data set is adopted, and SNR levels [-5, 0, 5, 10, 15, 20] dB are uniformly sampled \cite{MMSLLAMA2025}. The results for clean training condition are shown in Table \ref{tab:tab1}. From these results, we were unable to find significant difference or advantage for one fusion method over another. 
\begin{table}[tbh]
	\centering
	\caption{Baseline performance for clean training, WER (\%).}
	\begin{tabular}{|c||c||c||c||c|}
		\hline
		Fusion/SNR (dB) &SNR-5 & SNR0 & SNR5 & Clean \\	
		\hline	
		\hline		
		C\_Concat  & ${\bf 14.80}$ & $4.94$ & $1.71$ & $0.94$ \\		
		\hline	
		Add  & $15.55$ & ${4.26}$ & $1.55$ & ${\bf 0.80}$ \\						
		\hline
		Cross\_att & $17.04$ & $4.49$ & ${\bf 1.48}$ & $0.88$ \\
		\hline
		Q\_former & $16.09$ & $4.30$ & $1.62$ & $1.00$ \\
		\hline		
		\hline	
		T\_Concat & $18.53$ & $\bf 4.08$ & $1.74$ &$ 0.87$ \\
		\hline							
	\end{tabular}
	\label{tab:tab1}
\end{table}
The performance of noisy training is shown in Table \ref{tab:tab2}. Comparing Tables \ref{tab:tab1} and \ref{tab:tab2}, we can see that noisy training with data augmentation improved much in performance when the SNR levels for test conditions are low (e.g., SNR levels with 0 and -5 dBs) while there is no significant difference in clean test condition. From these results, we can see that channel-wise concatenation fusion is much better than other types of fusion methods in noisy training conditions. Therefore, in our later experiments, noisy training with channel-wise concatenation fusion methods is adopted.

\begin{table}[tbh]
	\centering
	\caption{Baseline performance for noisy training, SNR (dB), WER (\%).}
	\begin{tabular}{|c||c||c||c||c|}
		\hline
		Fusion/SNR (dB) &SNR-5 & SNR0 & SNR5 & Clean \\		
		\hline	
		\hline
		C\_Concat  & ${\bf 8.34}$ & ${\bf 2.82}$ & ${\bf 1.29}$ & $ 0.89$ \\		
		\hline	
		Add  & $9.06$ & $2.90$ & $1.45$ & ${\bf 0.85}$ \\						
		\hline
		Cross\_att & $10.38$ & $3.93$ & $1.40$ &$ 0.91$ \\
		\hline
		Q\_former & $9.97$ & $3.03$ &$ 1.44$ &$ 1.08$ \\
		\hline		
		\hline	
		T\_Concat & $11.41$ & $2.98$ & $1.48$ & $0.91$ \\
		\hline									
	\end{tabular}
	\label{tab:tab2}
\end{table}

\subsection{Effect of virtual bucket in OT}
By fixing the entropy regularization parameter $\lambda=0.1$ in Eqs. (\ref{eq:Qa2t}) and (\ref{eq:Qv2t}), and the alignment parameter $\alpha=0.1$ in Eq. (\ref{eq:TotalLoss}), we investigate the impact of incorporating a virtual bucket in the OT-based alignment framework during feature learning. Specifically, we compare model performance with and without the virtual bucket, and the results are presented in Table \ref{tab:tab3}. From Table \ref{tab:tab3}, we observe that introducing a virtual bucket in OT-based alignment yields a slight performance improvement. The virtual bucket plays a role similar to unbalanced OT by allowing unmatched mass to be absorbed \cite{PeyreBook2019}, thereby filtering out non-informative features or outliers during alignment. This mechanism is particularly meaningful in audio-visual speech recognition, where not all audio or visual frames necessarily correspond to useful linguistic information. Some frames may contain background noise, silence, or visually irrelevant content, and therefore should not be strictly aligned with linguistic tokens. Further investigation of virtual bucket modeling and its relationship to unbalanced OT will be considered in our future work, as we believe it provides a promising direction for improving robust multimodal alignment.
\begin{table}[tbh]
	\centering
	\caption{Performance without/with virtual bucket in OT for alignment,WER (\%).}
	\begin{tabular}{|c||c||c||c||c|}
		\hline
		Method/SNR (dB) &SNR-5 & SNR0 & SNR5 & Clean \\		
		\hline	
		\hline
		OT/wo bucket  &$ {8.06}$ &$ {2.61}$ & ${1.23}$ &  ${\bf 0.76}$	\\
		\hline	
		OT/w bucket  & ${\bf 7.98}$ & ${\bf 2.46}$ & ${\bf 1.15}$ &  $0.80$ \\ 					
		\hline								
	\end{tabular}
	\label{tab:tab3}
\end{table}

\subsection{Effect of alignment weight}
In loss function defined in Eq. (\ref{eq:TotalLoss}), setting different weighting coefficient $\alpha$ will emphasize coherent linguistic feature from audio and visual modalities during feature exploration. We do experiments to empirically check the effect with varying of $\alpha$, and show results in Table \ref{tab:tab4}. From this table, we can see that when $\alpha$ is in the range of $0.1$ to $0.5$, fairly well performance could be obtained.  
\begin{table}[tbh]
	\centering
	\caption{Performance of different weighting on alignment loss, WER (\%).}
	\begin{tabular}{|c||c||c||c||c|}
		\hline
		$\alpha$ /SNR (dB) &SNR-5 & SNR0 & SNR5 & Clean \\		
		\hline		
		\hline	
		$0.05$  & $8.12$ & $2.70$ &  $1.29$ &  $0.83$ \\		
		\hline	
		$0.1$  & $7.97$ & $2.41$ & $ 1.09$ &  $0.73$ \\		
		\hline			
		$0.3$  & $7.92$ & $\bf 2.36$ &  $\bf 1.08$ &  $\bf 0.71$ \\							
		\hline			
		$0.5$  & $\bf 7.86$ & $2.54$ &  $1.1$5 &  $0.88$ \\							
		\hline			
		$0.7$  & $8.82$ & $2.56$ &  $1.24$ &  $0.87$ \\							
		\hline		
		$0.9$  &$ 9.38$ & $2.76$ & $ 1.31$ &  $0.78$ \\							
		\hline		
	\end{tabular}
	\label{tab:tab4}
\end{table}





\subsection{Comparison with other methods}
We also have investigated the effects of several additional hyper-parameters in the proposed alignment framework, including the initial similarity margin $\tilde{s}$ for the virtual bucket in Eq. (\ref{eq:bucket}), the entropy regularization parameter $\lambda$ of OT in Eqs. (\ref{eq:Qa2t}) and (\ref{eq:Qv2t}), and the initial temperature scaling parameter $\tau$ in Eq. (\ref{eq:tempscale}). Our experiments indicate that these hyper-parameters exhibit strong interaction effects rather than independent influences on performance. In particular, multiple parameter combinations across relatively wide value ranges can yield similar performance, suggesting that the optimization landscape is nontrivial. Based on our current experiments, these hyper-parameters were selected empirically, and the resulting performances were compared with several advanced baseline methods, as shown in Table \ref{tab:tab5}. The results of our proposed methods are highlighted in gray with different parameter settings (setting 1 uses $\alpha=0.1$, $\tilde{s}=0.5$, $\lambda=0.2$, setting 2 uses $\alpha=0.3$, $\tilde{s}=0.5$, $\lambda=0.1$, and setting 3 uses $\alpha=0.1$, $\tilde{s}=0.5$, $\lambda=0.05$). From Table \ref{tab:tab5}, we observe that the proposed OT-based alignment framework for modality feature exploration consistently outperforms several advanced methods, even though the hyper-parameters have not yet been fully optimized. Moreover, explicitly setting modality alignment in feature exploration with reference to the token space of LLM significantly improved the performance (comparing our model settings with and without alignment module as shown in Fig. \ref{fig:fig1}). These results demonstrate both the effectiveness and robustness of the proposed approach. 

\begin{table}[tbh]
	\centering
	\caption{Comparison with other methods, WER (\%).}
	\begin{tabular}{|c||c||c||c||c|}
		\hline
		Methods /SNR (dB) &SNR-5 & SNR0 & SNR5 & Clean \\	
		\hline	
		\hline		
		Whisper-Flamingo (\cite{WFlamingo2024}) & $19.8$ &$5.0$ &$2.1$ &$1.5$ \\
		\hline		
		LLaMA-AVSR (\cite{CappellazzoICASSP2025} &$ 16.4$& $4.2$& $2.3$& $0.90$ \\ 
		\hline		
		MMS-LLaMA (\cite{MMSLLAMA2025}) &$ {7.44} $ & $2.66$ & $1.30$ & $0.95$ \\ 
		\hline		
		\hline	
		\rowcolor{lightgray}
		Our method (no align)  & $8.34$ &  $2.82$ & $1.29$ & $0.89$ \\		
		\hline	
		\rowcolor{lightgray}
		Our method (setting 1)  & $7.61$ &  $2.29$ & $1.09$ & $ {\bf 0.71}$ \\		
		\hline	
		\rowcolor{lightgray}
		Our method (setting 2)  & $7.96$ & $\bf 2.26$ & $\bf 1.07$ & $0.76$ \\	
		\hline	
		\rowcolor{lightgray}
		Our method (setting 3)  & ${\bf 7.16}$ & $ 2.34$ & $1.11$ & $0.73$ \\										
		\hline									
	\end{tabular}
	\label{tab:tab5}
\end{table}

\section{Conclusion and future work}
In this paper, we proposed an OT-based semantic alignment framework for LLM-AVSR. The proposed method addresses a fundamental limitation of existing LLM-AVSR systems, namely the representational discrepancy among acoustic, visual, and linguistic embedding spaces. Instead of directly fusing modality-specific features, we introduced an OT-based alignment module to establish structured correspondences between audio representations, visual representations, and the linguistic embedding space of the large language model. Specifically, acoustic features extracted by the Whisper encoder and visual features extracted by the AV-HuBERT encoder were aligned with reference to the token embedding space of the LLM. The OT-derived coupling matrices were further utilized as soft pseudo-labels to supervise contrastive learning, encouraging the model to learn semantically coherent and cross-modal consistent audio-visual representations. By explicitly aligning multimodal representations with linguistic semantics prior to fusion, the proposed framework improves multimodal feature integration and enables more effective conditioning of the LLM for speech recognition. Experimental results on the LRS3-TED benchmark demonstrated that the proposed framework consistently outperforms strong baseline systems under both clean and noisy conditions across various signal-to-noise ratio (SNR) settings. These results validate the effectiveness of OT-based semantic alignment for bridging heterogeneous multimodal representations and improving the robustness of LLM-AVSR.

The proposed alignment framework involves multiple hyper-parameters, including those related to OT estimation and contrastive learning. Determining the optimal combination of these hyper-parameters remains challenging, and in the current study they were selected empirically. Future work will investigate the sensitivity and interaction effects of these hyper-parameters through more systematic joint optimization to further improve model performance and training stability. 

\newpage

\section{Generative AI Use Disclosure}
We used generative AI tools for assistance in preparation of our paper. The AI tools were only employed for improving
and polishing in writing, readability, and formatting of the manuscript. They were not used to generate any substantial contents, core ideas, experimental design, analysis, or conclusions of the paper.

\end{document}